# Optical force on diseased blood cells: towards the optical sorting of biological matter

*Juan Sebastian Totero Gongora[1] and Andrea Fratalocchi[1,*]*

[1]*PRIMALIGHT, Faculty of Electrical Engineering, Applied Mathematics and Computational Science, King Abdullah University of Science and Technology (KAUST), Thuwal 23955-6900, Saudi Arabia*
[*]*Corresponding author. Email: andrea.fratalocchi@kaust.edu.sa - Tel: +966(12)8080348*

**Abstract**

By employing a series of massively parallel *ab-initio* simulations, we study how optical forces act on biological matter subject to morphological disease. As a representative case study, we here consider the case of *Plasmodium Falciparum* on red blood cells (RBC) illuminated by a monochromatic plane wave. Realistic parameters for the geometry and the refractive index are then taken from published experiments. In our theoretical campaign, we study the dependence of the optical force on the disease stage for different incident wavelengths. We show that optical forces change significantly with the disease, with amplitude variation in the hundreds of pN range. Our results open up new avenues for the design of new optical systems for the treatment of human disease.

# Introduction

In recent years, the interest in the study of the interaction between light and biological matter has largely increased. While optics has historically been mainly used for imaging applications, light has now become a tool of manipulation and direct interaction with biological samples. It is well known that when light impinges on an object it exerts a small force on it, whose values are in the range of tens of pN for micro sized specimens. Such optical force has two fundamental components: the gradient force, which is related to the electrostatic interaction between the polarization charges in the object and the field distribution itself, and the scattering force, due to the photon momentum transfer [1, 2]. The scattering part of the optical force depends on the wavenumber distribution of both the incident and scattered fields. Application examples of optical forces are found in optical tweezers, where a focused light beam is exploited to trap and transport micro/nano particles [3-6]. In biology, optical tweezers are usually employed as a non-invasive technique to manipulate cells and their sub-cellular components [7, 8]. While optical tweezers usually employ gradient forces, scattering forces have been recently investigated and a series of counterintuitive dynamics has been demonstrated, including fully optical tractor beams [9-11]. These techniques, including more pioneering idea such as e.g., cell optical sorting [12] and in-vivo manipulation [13] that are still under investigation, require a precise understanding on how the optical pressure affects biological organisms. This calls for the development of precise numerical methods, whose development will be crucial for applications of the aforementioned techniques. In this respect, *ab-initio* techniques represent a very important tool that can provide quantitative answers to the problem. Among the different approaches, the Finite Differences Time Domain (FDTD) method, yields a very flexible technique to study any experimental condition, as it is based on the numerical solution of Maxwell equations with no approximation [14]. Despite FDTD techniques have been applied in biology to simulate scattering from tissues and cells [15-17], nothing has been done in the direct computation of optical forces on bio samples.

In this work, we computed the optical forces acting on biological samples using FDTD simulations. In our FDTD approach, which explicitly considers the dispersive properties of the sample and of the environment, optical forces are expressed in terms of the Maxwell stress tensor formulation [18]. As an interesting case study, we measured the variation of optical force on a Red Blood cell when the morphology and the internal structure are changed, as in the case of RBC infected by *Plasmodium Falciparum*. Our results show that the optical force changes significantly with the morphology of the cell, allowing to sort RBC according to their disease progression.

This article is organized as follows. In section 1 we introduce the RBC model, describing its main properties as well as the typical deformation occurring in the different stages of a P. Falciparum infection. In section 2 we describe our FDTD calculation of optical forces within the Maxwell Stress Tensor formalism. In Section 3 we discuss the results of our simulations. Section 4 finally presents our conclusions.

# 1. Materials and Methods

In order to simulate a realistic biological sample, we considered Red Blood Cells, whose internal structure allows for a detailed optical modeling. Numerical simulations have shown that the presence of the cellular membrane does not provide significant contribution to the scattering properties of the RBC [19]. Moreover, in a healthy RBC, the cytoplasm of the cell is mainly homogeneous and it is primarily composed of Hemoglobin (Hgb) [20]. In terms of cell morphology, a healthy RBC exhibits a characteristic biconcave shape [21]. However, in the presence of an infection, the morphology and the internal structure of the RBC can change significantly. In some specific cases, such changes can be helpful in the diagnosis of the disease, as in the case of blood-strip based diagnosis of malaria [22]. In the past, much effort has been carried to model the optical response of RBCs in order to extract morphological information from scattering experiments and microscope measurements [19, 23, 24]. Among the different models proposed, the Evans-Fung model is widely used to represent the biconcave shape of a healthy RBC under different environmental conditions [25]. In the Evans-Fung model, which assumes an axisymmetric RBC, the profile of the cell can be represented by the thickness D(x) along the major axis (See Fig. 1a, red solid line), whose expression reads:

$$D(x) = \sqrt{1 - \left(\frac{x}{R_0}\right)^2} \left[C_0 - C_2 \left(\frac{x}{R_0}\right)^2 + C_4 \left(\frac{x}{R_0}\right)^4\right] \quad (1)$$

In Eq. (1) $R_0$ is the radius of the cell along the major axis, $C_0$ represents the central thickness and $C_2, C_4$ determine the biconcave shape curvature. These sets of parameters are usually extracted from scattering experiments or by directly fitting microscope or tomographic measurements of RBC samples [20, 26]. Typical values for a healthy RBC (hRBC) are

$R_0$=3.91 (µm), $C_0$=0.81 (µm), $C_2$=7.38 (µm), $C_4$=-4.39 (µm)

which correspond to an isotonic RBC [25].

*1.1 Optical modeling of Red Blood Cells infected by P. Falciparum*

As already mentioned, however, the presence of an infection can radically change the properties of the RBCs. In order to address a realistic case, we considered the modifications occurring in a red blood cell infected by *Plasmodium Falciparum*, one of the parasites responsible of malaria in humans. In the different stages of the parasite reproduction cycle, which is conventionally divided into ring (rRBC), trophozoite (tRBC) and schizont (sRBC) stages, the internal structure and the morphology of the infected RBC (iRBC) undergo radical changes (Fig. 1b). From a chemical point of view, the parasite consumes part of the Hgb from the RBC cytoplasm and produces high-refractive index Hemozoin crystals [27, 28]. From a morphological point of view, the iRBC loses its typical axisymmetric biconcave shape due to the growth of the parasite and, in the trophozoite and schizont stages, the cellular membrane of the iRBC becomes irregular due to the appearance of characteristic protuberances. It follows that a realistic optical description of the iRBC must include all the morphological changes at the different stages of infection, the presence of the parasite and the variation of concentration of Hgb in the cell cytoplasm. In our work, we modeled the irregular shape of an iRBC (Fig. 1a, blue dashed line) by applying a deformation displacement d**r** with definite statistical properties (Fig. 1a, yellow arrows) to every point of the curve describing a healthy RBC (Fig. 1a, solid red line) [29]. The expression of the deformation factor, which is directed along the normal to the surface, reads:

$$d\boldsymbol{r}(s) = G_\mu(s) \cdot A(s) \cdot \boldsymbol{n}(s), \quad (2)$$

where s is the curvilinear coordinate along the surface, **n**(s) is the normal unit vector, $G_\mu(s)$ is a Gaussian correlated random variable with mean value μ and correlation length $l_c$ and A(s) is an asymmetry multiplicative factor. The characteristic shape of the infected RBCs can be obtained by considering an asymmetry factor defined as:

$$A(s) = \exp\left[-\frac{[l(s) - L_0]^2}{2(L_2)^2}\right] - \Delta L \quad (3)$$

where l(s) represents the length of the curve as a function of the curvilinear coordinate s and $L_0, L_2$ and ΔL are length parameters which can be extracted from microscope or tomographic images of infected RBCs. As a result, Equations (2) and (3) provide an intuitive and physical model for the morphological changes occurring in an iRBC (Fig. 1b) and, due to the realization-dependent nature of the deformation factor, they allow for a statistical description of infected RBCs populations. In order to address the structural changes occurring in the interior of an iRBC, we modeled the parasite invading the iRBC as a rugged ellipsoid, whose geometrical parameters correspond to the experimental measurements reported in [30, 31]. As discussed in [32], the internal structure of the parasite has been divided into a cytoplasmic shell and a Hemozoin-rich core, with refractive index $n_{PF}(\lambda)$ and $n_H(\lambda)$, as functions of the wavelength, respectively [28]. The optical properties of Hgb as a function of its concentration have been subject of intensive study and several models have been proposed to describe the complex refractive index of a Hgb solution $ñ_{Hgb}(\lambda)$ as a function of wavelength. Following [27], we computed the complex refractive index for Hgb concentrations of 309g/l (healthy cell), 293g/l (rRBC), 233g/l (tRBC) and 187 g/l (sRBC) using the models and data included in [33-36].

## 2. Theoretical analysis and FDTD implementation

We compute optical forces with our massively parallel simulator NANOCPP, which is an homemade parallel c++ FDTD code that scales up to hundreds of thousands of processors [37]. In the FDTD algorithm, Maxwell's equations are solved simultaneously in space and time, on a lattice specifically designed to enforce divergence conditions and within a second order time accurate algorithm [14]. FDTD provides a full knowledge over all the field components, both in the time and space domain, with arbitrary precision. In order to reduce the computational time, we resorted to parallel computing techniques and we decompose the spatial domain across a large number of processors, which solve Maxwell's equations in parallel. In our implementation, all the exchange of information among the processors is based on the Message-Passage-Interface (MPI) standard. Electromagnetic sources are then included using an exact implementation based on the Total-Field-Scattered-Field (TFSF) formulation, including fully customizable Uniaxial-Perfectly-Matched-Layer (UPML) to simulate realistic open systems [14].

We compute the optical force in terms of the Maxwell Stress Tensor $(\overleftrightarrow{T})_{ij}$, defined as:

$$(\overleftrightarrow{T})_{ij} \stackrel{\text{def}}{=} \varepsilon_0 \left(E_i E_j - \frac{1}{2}\delta_{ij}E^2\right) + \mu_0 \left(H_i H_j - \frac{1}{2}\delta_{ij}H^2\right) \, , \quad i = x, y, z. \quad (4)$$

where $\varepsilon_0$ is the vacuum permittivity, $\mu_0$ is the vacuum permeability, $\delta_{ij}$ is the Kronecker delta and $\mathbf{E}=(E_x,E_y,E_z)$ and $\mathbf{H}=(H_x,H_y,H_z)$ are the electric and magnetic fields, respectively [18]. Given Eq. (4), the optical force $\mathbf{F}=(F_x,F_y,F_z)$ exerted by an electromagnetic wave impinging on an object is then obtained from the time average of the flux of the Maxwell Stress Tensor over an arbitrary enclosing surface $S$, namely

$$\vec{F} = \langle \oiint_S \overleftrightarrow{T} \cdot d\vec{S} \rangle_T \quad (5)$$

$$F_i = \langle \sum_j \oiint_S T_{ij} dS_j \rangle_T \, , \quad i = x, y, z \, , \quad (6)$$

Where the time average is defined over an optical cycle T as follows:

$$\langle f(t) \rangle_T = \frac{1}{T} \int_0^T f(t) \, . \quad (5)$$

The field-dependent nature of $(\overleftrightarrow{T})_{ij}$ simplifies the computation of the force $F_i$ in a FDTD simulation, where the electric and magnetic fields are available in every point of the discretized domain. Equations (3)-(5) are computed by a trapezoidal integration of the field components along the surface $S$, and by time averaging the resulting values. With this implementation, optical forces acting on a generic system can be computed with an arbitrary precision, limited in principle only by the available computational resources.

## 3. Results and Discussion

Figure 2a summarizes the FDTD setup used in our numerical experiments: a two-dimensional RBC is placed inside a 9μm×4μm TFSF region, where a monochromatic plane wave of wavelength λ propagates along the $\hat{\mathbf{z}}$ direction. The medium surrounding the RBC we consider dispersive plasma with refractive index $n_m(\lambda)$ [38]. The incident power density is fixed at 8.4mW/μm², which matches the actual value used in *in-vivo* experiments with optical tweezers and therefore lies below the radiation damage threshold of RBCs [13]. A typical spatial distribution of the total electromagnetic energy is displayed in Fig. 2b, where a healthy RBC is illuminated with a λ=480nm plane wave. To eliminate any artificial boundary effect in the calculation of the optical force, we employed a thick (60 points) UPML layer. We considered a spatial domain corresponding to a volume of 10μm×5μm, with a discretization of 2500×1250 points, equivalent to a resolution of dx=dz=4nm. This corresponds to a resolution between 65 and 286 points per internal wavelength, which ensures a very high-accuracy in the force computation. In our simulations, we considered wavelengths in the both visible and IR regions, which can penetrate the tissues. The integration surface $S$ is symmetrically placed 100nm apart from the RBC surface, sufficiently near to capture the near-field features of the scattered field. In our simulations, we considered realistic dispersion parameters of the Hemoglobin, as taken from the literature [33-36]. Figure 3 reports the behavior of the refractive index $n_{HGB}$ and the absorption $k_{HGB}$ for different wavelengths and for different concentrations. As seen from Fig. 3, both quantities $n_{HGB}$ and $k_{HGB}$ vary strongly as a function of the wavelength and the concentration.

In our FDTD experiments, we measured the component of the force in the propagation direction ($F_z$). In order to investigate the dependence of the optical force to the geometrical parameters, we performed an extensive massively parallel campaign, composed of approx. 5M of CPU hours on 4096 processors. In each simulation, we change both the frequency and the morphology of the RBC, according to the different stages of the disease illustrated in Fig. 1b. Figure 4 summarizes our results. Quite interestingly, the force manifests a significant change when the disease progressively increases, showing characteristic resonances whose position and peak-to-peak amplitude is strongly wavelength dependent. For each frequency, the optical forces manifests a monotonic decrease from the healthy stage (hRBG) to the diseased case (sRBC). This change is quite dramatic at the specific wavelengths where the hemoglobin is strongly absorbing, showing that the optical properties of the molecule are the dominant factor in the determination of the force. At the wavelength of 425nm, the optical force manifest an impressive change from approximately 120pN for hRBC to 40pN in the case of sRBC, with a reduction of 300%.

These results can be relevant in the design and realization of new types of microfluidic channels [39-42] for the diagnosis and treatment of blood disease, such as malaria, where healthy blood cells are all optically selected and diseased cells are filtered out, opening new perspectives in optical nanomedicine.

## 4. Conclusions

In this work we presented an *ab-initio* analysis on the optical forces acting on biological objects. As an interesting case study, we considered red blood cells subjected to malaria disease. Optical forces are calculated from the Maxwell stress tensor formalism under a massively parallel FDTD campaign. Our results show that, even in the absence of any focused beam (as those used in trapping or tweezing experiments) the force is highly sensitive to changes in the cell morphology and sufficiently high (in the hundreds of pN) range to sort the particles according to their ill conditions. This opens to the development of new techniques for the *in-vivo* treatment of diseases that affect the morphology of biological matter.

# Figures

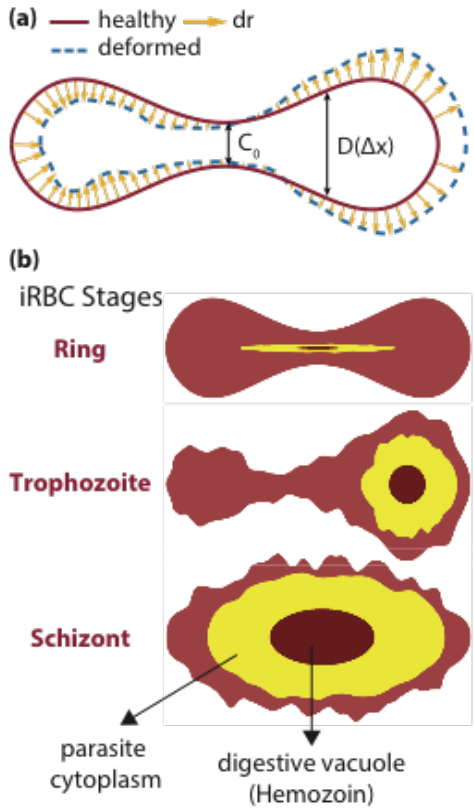

**Figure 1. Analytic representation of healthy and *P. Falciparum infected* RBC.** (a) In the Evans-Fung model, the profile of a healthy RBC (solid red line) can be represented in terms of the cell thickness D(x) (Eq. 1). In the presence of an infection, conversely, the profile of the cell undergoes complex morphological changes and the deformed cell (blue dashed line) can be modeled in terms of a deformation factor **dr** (yellow arrows) which takes into account both asymmetry and membrane fluctuations (Eqs. 2 and 3). (b) By combining Eqs. 1-3, we can model the three phenotypes of iRBC occurring in the presence of a *P. Falciparum* infection. In addition to the variation of the cell profile, we included the structural changes occurring in the interior of the RBC due to the growth of the parasite.

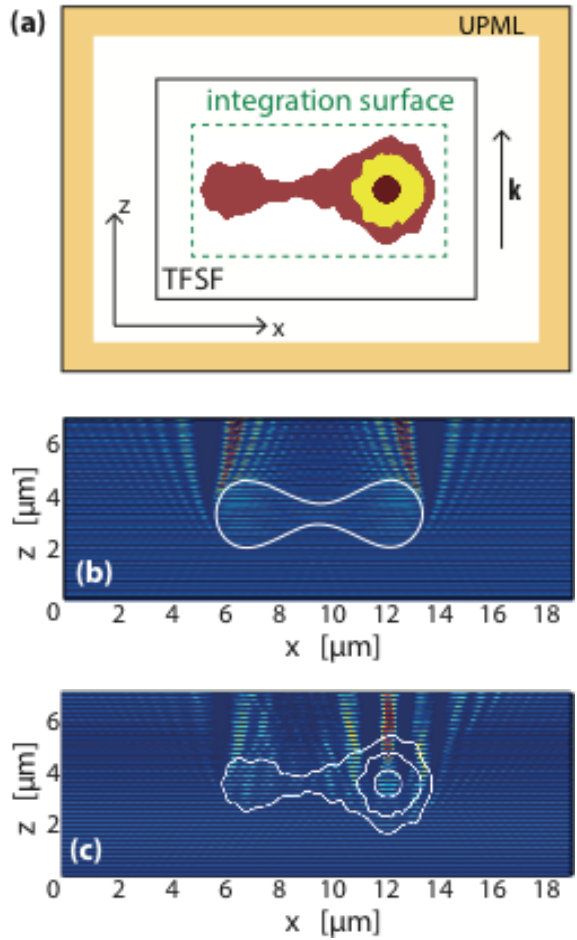

**Figure 2. FDTD Setup.** (a) FDTD setup: we enclosed the RBC within a rectangular TFSF region (solid black line). The force $F_z$ is measured along the propagation direction by integrating the Maxwell Stress Tensor (Eq. 4) over a surface (green dashed line) enclosing the cell. (b-c) Spatial electromagnetic energy distribution at the wavelength λ=460nm in the case of (b) a healthy cell and (c) a trophozoite iRBC. As can be evinced from the figure, in the later stages of the infection the digestive vacuoles of the parasite, rich in Hemozoin, produce a completely different scattering pattern.

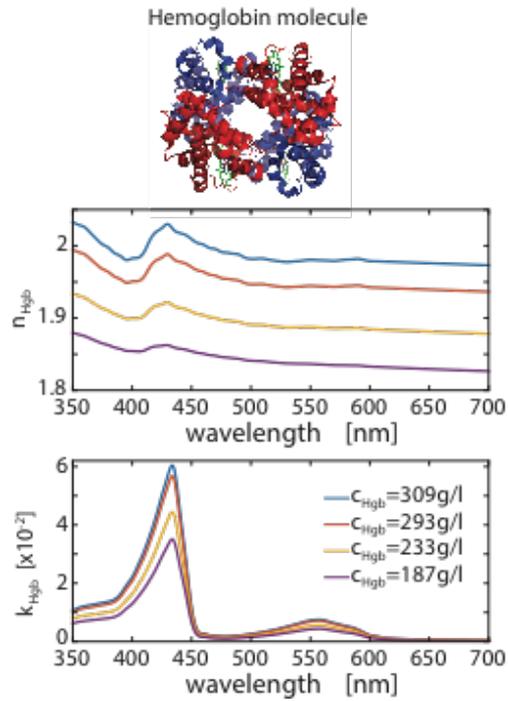

**Figure 3. Optical properties of iRBC cytoplasm.** (a) Real and (b) imaginary parts of the Hemoglobin refractive index as a function of wavelength and Hgb Concentration. Across the different stages of infection, the dispersion and absorption properties of the RBC are linked to the decrease of Hgb concentration in the cytoplasm. In our simulations we considered complex a refractive index corresponding to Hgb concentrations of 309g/l (healthy cell), 293g/l (rRBC), 233g/l (tRBC) and 187 g/l (sRBC) [27].

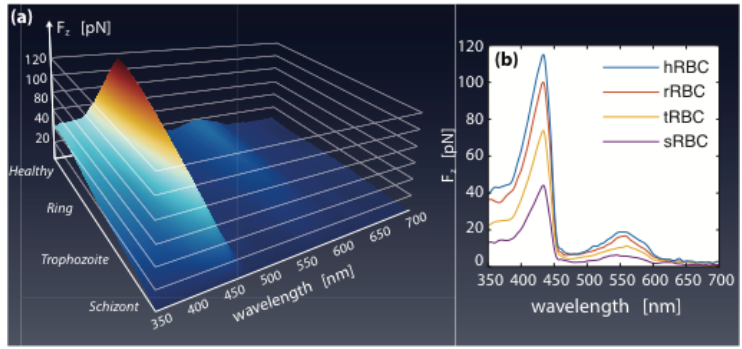

**Figure 4. Optical forces on iRBC at different stages of P. Falciparum infection: FDTD Results.** (a) Surface plot of the optical force $F_z$ as a function of wavelength and disease progression. Due to the strong decrease in Hgb concentration, the optical force exhibits up to a 300% variation between a healthy RBC (b, blue solid line) and an iRBC in an advanced infection stage (b, violet line).